\documentstyle[12pt,bezier,epsfig]{article}
\begin{document}
\def \inbar{\vrule height1.5ex width.4pt depth0pt}
\def \xC{\relax\hbox{\kern.25em$\inbar\kern-.3em{\rm C}$}}
\def \xR{\relax{\rm I\kern-.18em R}}
\newcommand{\xZ}{Z \hspace{-.08in}Z}
\newcommand{\xbe}{\begin{equation}}
\newcommand{\xee}{\end{equation}}
\newcommand{\xbea}{\begin{eqnarray}}
\newcommand{\xeea}{\end{eqnarray}}
\newcommand{\xnn}{\nonumber}
\newcommand{\xkt}{\rangle}
\newcommand{\xbr}{\langle}
\newcommand{\xlll}{\left( }
\newcommand{\xrrr}{\right)}
\newcommand{\xcun}{\mbox{\footnotesize${\cal N}$}}

\title{Cosmological Adiabatic Geometric Phase of a Scalar Field in a Bianchi Spacetime\footnote{Presented in the IX Regional Conference on Mathematical Physics, held in the Feza Gursey Institute, Istanbul, August 1999.}}
\author{Ali Mostafazadeh\\ \\
Department of Mathematics, Ko\c{c} University,\\
Rumelifeneri Yolu, 80910 Sariyer, Istanbul, Turkey\\
amostafazadeh@ku.edu.tr}
\date{}
\maketitle

\begin{abstract}
A two-component formulation of the Klein-Gordon equation is used
to investigate the cyclic and noncyclic adiabatic geometric phases 
due to spatially homogeneous (Bianchi) cosmological models. It is 
shown that no adiabatic geometric phases arise for Bianchi type I 
models. For general Bianchi type IX models the problem of the 
adiabatic geometric phase is shown to be equivalent to the one for
nuclear quadrupole interactions of a spin. For these models 
nontrivial non-Abelian adiabatic geometrical phases may occur
in general. 
\end{abstract}

\baselineskip=24pt

\section*{I~~Introduction}

In Ref.~\cite{p19a} a two-component formulation of the Klein-Gordon
equation is used to develop relativistic analogues of the quantum
adiabatic approximation and the adiabatic dynamical and geometric
phases. This method provides a precise definition of an adiabatic
evolution of a Klein-Gordon field in a curved background spacetime.
The purpose of this article is to employ the results of
Ref.~\cite{p19a} in the investigation of geometric phases due to
a spatially homogeneous background spacetime.

The phenomenon of geometric phase in gravitational systems has been
previously considered by Cai and Papini \cite{ca-pa}, Brout and
Venturi \cite{br-ve}, Venturi \cite{ve1,ve2}, Casadio and Venturi
\cite{ca-ve}, Datta \cite{da}, and Corichi and Pierri \cite{co-pi}.
The motivation for these studies extends from the investigation of
the study of weak gravitational fields \cite{ca-pa} to various
problems in quantum cosmology and quantum gravity 
\cite{br-ve,ve1,ve2,ca-ve,da}. 

These studies were mostly plagued by the problem of constructing
appropriate inner products on the space of solutions of the Klein-Gordon
equation. This problem has so far been solved for stationary spacetimes.
Therefore the existing results have a limited domain of applicability.
The most important feature of the method developed in Ref.~\cite{p19a}
is that it avoids the above problem by showing that indeed the adiabatic
geometric phase is independent of the choice of an inner product on the
space of solutions of the field equation. This enables one to investigate
the phenomenon of the adiabatic geometric phase for spatially homogeneous
spacetimes which are clearly non-stationary. One must, however, note that
the results presented in this article are relevant to adiabatically
evolving Klein-Gordon fields \cite{p19a}. This raises the question
whether the adiabaticity of the evolution is compatible with the fact
that the background spacetime is non-stationary. It turns out that the
answer to this question is in the positive, i.e., there are 
Klein-Gordon fields in a non-stationary spacetime which do have
adiabatic evolutions.

The organization of the paper is as follows. In section~II, the results of 
Ref.~\cite{p19a} which will be used in this paper are briefly reviewed. 
The computation of the geometric phase for general spatially homogeneous 
(Bianchi) models are discussed in  Section~III. These are applied in the
analysis of the geometric phase problem for Bianchi type~I and type~IX
models in  sections~IV and~V, respectively. A summary of the main results
and the concluding remarks are given in section~VI.

Throughout this paper the signature of the spacetime metric g is taken to
be $(-,+,+,+)$. Letters from the beginning and the middle of the Greek
alphabet are associated with an arbitrary local basis and a local
coordinate basis of the tangent spaces of the spacetime manifold,
respectively. The letters from the beginning and the middle of the 
Latin alphabet label the corresponding spatial components and take 
values in $\{1,2,3\}$.

\section*{II~~Two-Component Formalism and the Adiabatic Geometric 
Phase}

Consider a complex scalar field $\Phi$ defined on a globally 
hyperbolic spacetime $(M,{\rm g})=(\xR\times\Sigma,{\rm g})$
satisfying
	\xbe
	\left(g^{\mu\nu}\nabla_\mu\nabla_\nu-\mu^2\right)\Phi=0\;,
	\label{k-g-1}
	\xee
where $g^{\mu\nu}$ are components of the inverse of the metric 
${\rm g}$, $\nabla_\mu$ is the covariant derivative along 
$\partial/\partial x^\mu$ defined by the Levi Civita connection, and 
$\mu$ is the mass. 

Denoting a time derivative by a dot, one can express
Eq.~(\ref{k-g-1}) in the form
	\xbe
	\ddot\Phi+\hat D_1\dot\Phi+\hat D_2\Phi=0\;,
	\label{fi-eq}
	\xee
where
	\xbea
	\hat D_1&:=&\frac{1}{g^{00}}\left[ 2 g^{0i}\partial_i-
	g^{\mu\nu}\Gamma_{\mu\nu}^0\right]\;,
	\label{d1}\\
	\hat D_2&:=&\frac{1}{g^{00}}\left[ g^{ij}
	\partial_i\partial_j-g^{\mu\nu}
	\Gamma_{\mu\nu}^i\partial_i-\mu^2\right]\,.
	\label{d2}
	\xeea

A two-component  representation of the field equation (\ref{fi-eq})
is
	\xbe
	i\dot\Psi^{(q)}=\hat H^{(q)}\Psi^{(q)}\;,
	\label{sch-eq-q}
	\xee
where
	\xbea
	\Psi^{(q)}&:=&\left(\begin{array}{c} u^{(q)}\\v^{(q)}
	\end{array}\right)\,,
	\label{psi-q}\\
	u^{(q)}&:=&\frac{1}{\sqrt{2}}\:(\Phi+q\dot\Phi)\;,~~~
	v^{(q)}\::=\:\frac{1}{\sqrt{2}}\:(\Phi-q\dot\Phi)\;,	
	\label{u-v-q}\\
	\hat H^{(q)}&:=&\frac{i}{2}\left(
	\begin{array}{cc}
	\frac{\dot q}{q}+\frac{1}{q}-\hat D_1-q\hat D_2&
	-\frac{\dot q}{q}-\frac{1}{q}+\hat D_1-q\hat D_2\\
	&\\
	-\frac{\dot q}{q}+\frac{1}{q}+\hat D_1+q\hat D_2&
	\frac{\dot q}{q}-\frac{1}{q}-\hat D_1+q\hat D_2
	\end{array}\right)\,,
	\label{h-q}
	\xeea
and $q$ is an arbitrary, possibly time-dependent, nonzero complex
parameter. 

Next consider the eigenvalue problem for $H^{(q)}$. Denoting the
eigenvalues and eigenvectors by $E_n^{(q)}$ and $\Psi_n^{(q)}$,
i.e.,
	\xbe
	H^{(q)}\Psi_n^{(q)}=E_n^{(q)}\Psi_n^{(q)}\;,
	\label{eg-va-eq-q}
	\xee
one has \cite{p19a}
	\xbe
	\Psi_n^{(q)}=\frac{1}{\sqrt{2}}\:\left(\begin{array}{c}
	1-iqE_n^{(q)}\\
	1+iqE_n^{(q)}
	\end{array}\right)\: \Phi_n^{(q)}\;,
	\label{eg-ve}
	\xee
where $\Phi_n^{(q)}$ satisfies
	\xbe
	\left[ \hat D_2-iE_n^{(q)}(\hat D_1-\frac{\dot q}{q})
	-\left( E_n^{(q)}\right)^2\right] \Phi_n^{(q)}=0\;.
	\label{ge-eg-va-eq}
	\xee
This equation defines both the vectors $\Phi_n^{(q)}$ and the complex 
numbers $E_n^{(q)}$. 

The following is a summary of some of the results obtained in 
Ref.~\cite{p19a}.
	\begin{itemize}
	\item[1)] Eq.~(\ref{ge-eg-va-eq}) reduces to the ordinary
eigenvalue equation for $\hat D_2$, if $\hat D_1=\dot q/q$. In this
case $\Phi_n^{(q)}$ and $E_n^{(q)}$ are independent of the choice
of $q$, and one can drop the labels $(q)$ on the right hand side
of  Eq.~(\ref{eg-ve}). 
	\item[2)] $\Phi_n^{(q)}$ belong to the Hilbert space 
${\cal H}_t=L^2(\Sigma_t)$ of square-integrable functions on the spacelike
hypersurfaces $\Sigma_t$ where the integration is defined by the measure 
$[\det(^{(3)}{\rm g})]^{1/2}$ and $^{(3)}{\rm g}$ is the Riemannian 
three-metric on $\Sigma_t$ induced by the four-metric ${\rm g}$.
	\item[3)] Suppose that 
	\begin{itemize}
	\item[3.1)]  $\hat D_1=\dot q/q$,
	\item[3.2)]  $\hat D_2$ is self-adjoint with respect to the 
	inner product $\xbr~,~\xkt$ of ${\cal H}_t$,
	\item[3.3)]  $\hat D_2$ has a discrete spectrum,
	\item[3.4)]  during the evolution of the system $E_n\neq
	E_m$ iff $m\neq n$, i.e., there is no level-crossings, and
	\item[3.5)]  $E_n$ is a non-degenerate eigenvalue of $H^{(q)}$,
	alternatively $E_n^2$ is a non-degenerate eigenvalue of 
	$\hat D_2$.
	\end{itemize}
Then for all $m\neq n$, $\xbr\Phi_m|\dot\Phi_n\xkt=
\xbr\Phi_m|\dot{\hat D}_2\Phi_n\xkt/(E_n^2-E_m^2)$. The
relativistic adiabatic approximation corresponds to the case
where the latter may be neglected. In this case, an initial
two-component Klein-Gordon field
	\xbe
	\Psi^{(q)}(0)= e^{i\alpha_n(0)}\Psi_n^{(q)}(0)
	+e^{i\alpha_{-n}(0)}\Psi_{-n}^{(q)}(0)\;,
	\label{ini-re-ad-ev}
	\xee
with $n\geq 0$ and $\alpha_{\pm n}(0)\in\xC$, evolves
according to
	\xbe
	\Psi^{(q)}(t)\approx e^{i\alpha_n(t)}\Psi_n^{(q)}(t)
	+e^{i\alpha_{-n}(t)}\Psi_{-n}^{(q)}(t)\;,
	\label{re-ad-ev}
	\xee
where $\alpha_{\pm n}(t)=[\alpha_{n}(0)+\alpha_{- n}(0)]/2
+\gamma_n(t)+\delta_{\pm n}(t)$, $\alpha_{\pm n}(0)$ are arbitrary
constants, $\gamma_n(t)$ is the geometric part of both $\alpha_{\pm n}(t)$
and $\delta_{\pm n}(t)$ is the dynamical part of $\alpha_{\pm n}(t)$. 
They are given by
	\xbea
	\gamma_n(t)&=&\int_{R(0)}^{R(t)}{\cal A}_n[R]\;,
	\label{x1}\\
	\delta_{\pm}(t)&=&i\xi_n(t)\pm\frac{\eta_n(t)}{2}\;,
	\label{x3}
	\xeea
where
	\xbe
	{\cal A}_n[R]:=\frac{i\xbr\Phi_n[R]|d|\Phi_n[R]\xkt}{
	\xbr\Phi_n[R]|\Phi_n[R]\xkt}=
	\frac{i\xbr\Phi_n[R]|\frac{\partial}{\partial R^a}|
	\Phi_n[R]\xkt}{
	\xbr\Phi_n[R]|\Phi_n[R]\xkt}\,dR^a\;,
	\label{be-co}
	\xee
is the Berry's connection one-form \cite{berry1984}, $R=(R^1,\cdots,R^n)$
are the parameters of the system, i.e., the components of the metric,
$d$ stands for the exterior derivative with respect to $R^a$,
	\xbea
	\xi_n(t)&:=&\frac{1}{2}\int_0^t f_n(t') [1-\cos\eta_n(t')]dt'\;,
	\label{x4}\\
	f_n(t)&:=&\frac{d}{dt}\ln [q(t) E_n(t)]\;,
	\label{x5}
	\xeea
and $\eta_n$ is the solution of
	\xbe
	\dot\eta_n+f_n\sin\eta_n+2E_n=0\;,~~~~
	\eta_n(0)=\alpha_n(0)-\alpha_{-n}(0)\;.
	\label{x6}
	\xee
In view of Eqs.~(6), (7), and (10), Eq.~(\ref{re-ad-ev}) can be written
in the form
	\xbe
	\Phi=c\, e^{i\gamma_n(t)}(e^{i\delta_n(t)}+e^{i\delta_{-n}(t)})
	\Phi_n,
	\label{zzz}
	\xee
where $\Phi$ is the one-component Klein-Gordon field, i.e., the solution
of the original Klein-Gordon equation~(1), and $c:=\exp[\alpha_n(0)+
\alpha_{-n}(0)]$.
	\item[4)] If in addition to the above adiabaticity
condition one also has $\dot E_n\approx 0$, then an initial 
eigenvector $\Psi_n^{(q)}(0)$ evolves according to
	\[\Psi^{(q)}(t) \approx e^{i\alpha_n(t)}\Psi_n^{(q)}(0)\;,\]
where the total phase angle $\alpha_n$ again consists of a geometric
and a dynamical part. This case corresponds to what is termed as 
ultra-adiabatic evolution in Ref.~\cite{p19a}.
	\end{itemize}
	Suppose that conditions 3.1) -- 3.4)
are satisfied, but $E_n$ is ${\cal N}$ fold degenerate. Then the condition
for the validity of the adiabatic approximation becomes 
$\xbr\Phi_m^I|\dot\Phi_n^J\xkt\approx 0$, for all $m\neq n$ and 
$I,J=1,2,\cdots {\cal N}$. Here $\Phi_n^1,\cdots \Phi_n^{\cal N}$ are 
orthogonal eigenvectors spanning the degeneracy subspace of ${\cal H}_t$ 
associated with $E_n^2$. In this case, $\exp[i\alpha_{\pm n}(t)]$ become 
matrices of the form $e^{i\delta_{\pm n}(t)}\Gamma_n(t)$ where
	\xbe
	\Gamma_n(t):={\cal P}\, \exp[{i\int_{R(0)}^{R(t)}{\cal A}_n}]\;,
	\label{e^ia}
	\xee
${\cal P}$ is the path-ordering operator, ${\cal A}_n$ is a matrix of one-forms 
with components
	\xbe
	{\cal A}_n^{IJ}[R]:=
	\frac{i \xbr \Phi^I_n[R]|d| \Phi^J_n[R]\xkt}{\xbr\Phi^I_n[R]|
	\Phi^I_n[R]\xkt}\;.
	\label{connection}
	\xee
In this case, a solution of the one-component Klein-Gordon equation~(1) 
is given by
	\xbe
	\Phi=\sum_{I,J=1}^N(c_{n}^I e^{i\delta_n(t)}+c_{-n}^I 
	e^{i\delta_{-n}(t)})\Gamma_n^{IJ}(t)\Phi_n^J\;,
	\label{www}
	\xee 
where $c_{\pm n}^I$ are constant coefficients determined by the initial 
conditions. 

If the parameters $R$ undergo a
periodic change, i.e., $R(T)=R(0)$ for some $T\in\xR^+$, then the 
path-ordered exponential $\Gamma_n(T)$ which is called the {\em cyclic 
adiabatic geometrical phase},  cannot be removed by a gauge 
transformation. Eq.~(\ref{connection}) shows that the formula for the 
relativistic adiabatic geometric phase has the same form as its non-relativistic 
analogue \cite{wi-ze}. In particular for ${\cal N}=1$ (the non-degenerate case), 
one recovers Berry's connection one-form (\ref{be-co}).
In this case $\Gamma_n(T)$ reduces to an ordinary exponential and yields
the Berry phase $e^{i\gamma(T)}$ for the Klein-Gordon field.

In the remainder of this paper, I shall consider the problem of the
adiabatic geometric phase for a Klein-Gordon field minimally
coupled to a spatially homogeneous gravitational field.

It is important to note that the cyclic geometric phase has physical
significance, if one has a cyclic evolution\footnote{For a discussion of the 
meaning of a cyclic evolution of a Klein-Gordon field see 
Ref.~\cite{p19a}.}. For an adiabatic evolution, the evolving state 
undergoes a cyclic evolution, if the parameters of the system, in this case
the components of the metric tensor, change periodically in time. This
corresponds to the spatially homogeneous (Bianchi) cosmological models
which admit periodic (oscillatory) solutions\footnote{For an example see 
Ref.~\cite{ryan}.}. This observation does not however mean that the 
connection one-forms ${\cal A}_n$ and their path-ordered exponentials
$\Gamma_n(t)$ lack physical significance for general nonperiodic Bianchi 
models. The cyclic adiabatic geometric phases have noncyclic analogues 
which occur in the evolution of any quantum state, \cite{noncyclic,p31}. 

A noncyclic analogue of the non-Abelian cyclic geometric phase has 
recently been introduced by the present author \cite{p31}. In view of
the results of Ref.~\cite{p31}, the {\em noncyclic adiabatic geometric 
phase} for an adiabatically evolving Klein-Gordon field is given by
	\xbe
	\tilde\Gamma_n(t):=w_n(t)\Gamma_n(t)\;,
	\label{noncyc-G}
	\xee
where $w_n(t)$ is an $\xcun\times\xcun$ matrix with entries:
	\xbe
	w_n^{IJ}(t):=\xbr\Phi_n^I[R(0)]|\Phi^J_n[R(t)]\xkt\;.
	\label{noncyc-w}
	\xee
The noncyclic geometric phase has the same gauge transformation 
properties as the cyclic geometric phase. In particular, its eigenvalues
and its trace are gauge-invariant physical quantities, \cite{p31}.

If it happens that the connection one-form ${\cal A}_n$ is exact, i.e., it
is a pure gauge, then there are two possibilities:
	\begin{itemize}
	\item[a)] The curve $C$ traced by the parameters $R$ of the system
in time has a part which is a non-contractible loop. In this case the cyclic 
or noncyclic geometric phase  will be a topological
quantity analogous to the Aharonov-Bohm phase \cite{berry1984}. Such
a geometric phase will be called a {\em topological phase}. Topological 
phases can occur only if the parameter space of the system has a nontrivial
first homology group.
	\item[b)] The curve $C$ does not have a piece which is a 
non-contractible loop. In this case, the geometric phase is either unity (the
cyclic case) or it depends only on the end points, $R(0)$ and $R(t)$, of 
$C$ (the noncyclic case). Such a geometric phase will be called a {\em 
trivial geometric phase}.
	\end{itemize}

\section*{III~~Spatially Homogeneous Cosmological Models}
Consider  Klein-Gordon fields in a spatially homogeneous (Bianchi)
cosmological background associated with a Lie group
$G$, i.e., $M=\xR\times G$. In a synchronous invariant
basis the spacetime metric g is given by its spatial
components $g_{ab}$:
	\xbe
	ds^2=g_{\alpha\beta}\omega^\alpha\omega^\beta=
	-dt^2+g_{ab}\omega^a\omega^b\;,
	\label{ds2}
	\xee
where $\omega^a$ are the left invariant one-forms
and $g_{ab}=g_{ab}(t)$.  Throughout this article I use the
conventions of Ref.~\cite{ra-sh}.

The first step in the study of the phenomenon of the adiabatic 
geometric phase due to a spatially homogeneous cosmological background
is to compute the operators $\hat D_1$ and $\hat D_2$ of
Eqs.~(\ref{d1}) and (\ref{d2}) in the invariant basis.  It is not
difficult to see that with some care these equations are valid
in any  basis. One must only replace the coordinate
labels ($\mu,\nu,\cdots, i,j,\cdots$) with the basis (in this case
invariant basis) labels $(\alpha,\beta,\cdots,a,b,\cdots)$,
and interpret $\partial_a$ as the action of the operators
$\hat X_a$ associated with the dual vector fields to $\omega^a$.
This leads to
	\xbea
	\hat D_1&=&g^{ab}\Gamma_{ab}^0\;,
	\label{d1-bc}\\
	\hat D_2&=& -\Delta_t+\mu^2\;,
	\label{d2-bc}\\
	\Delta_t&:=&g^{ab}\nabla_a\nabla_b\:=\:
	g^{ab}\hat X_a\hat X_b-\Gamma_{ab}^c\hat X_c\;,
	\label{laplacian-bc}
	\xeea
where $\Delta_t$ is the Laplacian on $\Sigma_t$, $\nabla_a$
are the covariant derivatives corresponding to the Levi Civita
connection,
	\xbe
	\Gamma_{\alpha\beta}^\gamma:=\frac{1}{2}\,
	g^{\gamma\delta}(g_{\delta\alpha,\beta}+
	g_{\beta\delta,\alpha}-g_{\alpha\beta,\delta}+
	g_{\epsilon\alpha}C^{\epsilon}_{\delta\beta}+
	g_{\epsilon\beta}C_{\delta\alpha}^\epsilon)-
	\frac{1}{2}\:C_{\alpha\beta}^\gamma\;,
	\label{gamma}
	\xee
as derived in Ref.~\cite{ra-sh}, $g_{\alpha\beta,\gamma}:=
\hat X_\gamma g_{\alpha\beta}$, and  $C_{\alpha\beta}^\gamma$ are 
the structure constants:
	\xbe
	\left[\hat X_\alpha,\hat X_\beta\right]=-
	C_{\alpha\beta}^\gamma\, \hat X_\gamma\;,
	\label{lie-al}
	\xee
with $\hat X_0:=\partial/\partial t$. In view of the latter
equality, the structure constants with a time label vanish.
This simplifies the calculations of $\Gamma$'s. The only
nonvanishing ones are
	\xbea
	\Gamma_{ab}^0&=&\frac{1}{2}\:\dot g_{ab}\;,
	\label{gamma-ab0}\\
	\Gamma_{ab}^c&=&\frac{1}{2}\: g^{cd}(g_{ea}C^e_{db}+
	g_{eb}C^e_{da})-\frac{1}{2}\:C_{ab}^c\;.
	\label{gamma-abc}
	\xeea
In view of these relations, the expression for $\hat D_1$
and $\hat D_2$ may be further simplified:
	\xbea
	\hat D_1&=& \frac{\partial}{\partial t}\ln\sqrt{g}\;,
	\label{d1-bc-2}\\
	\hat D_2&=& -\Delta_t+\mu^2\:=\:-(
	g^{ab}\hat X_a\hat X_b-C_{ab}^bg^{ac}\hat X_c)+\mu^2\;,
	\label{d2-bc-2}
	\xeea
where $g$ is the determinant of $(g_{ab})$. Note that for
a unimodular, in particular semisimple, group $C_{ab}^b=0$,
and the second term on the right hand side of  (\ref{d2-bc-2})
vanishes. The corresponding Bianchi models are knows as
Class A models.

As seen from Eq.~(\ref{d1-bc-2}), $\hat D_1$ acts by
multiplication by a time-dependent function. Therefore,
choosing $q=i\sqrt{g}$, $\hat D_1=\dot q/q$. This reduces 
Eq.~(\ref{ge-eg-va-eq}) to the eigenvalue equation
	\xbe
	(\hat D_2-E_n^2)\Phi_n=-(\Delta_t+E_n^2-\mu^2)
	\Phi_n=0\;,
	\label{eg-va-eq-la}
	\xee
for the operator $\hat D_2$ which being essentially the
Laplacian over $\Sigma_t$, is self-adjoint. This
guarantees the orthogonality of $\Phi_n$ and the reality of
$E_n^2$. \footnote{One can also show that, since $q$ is 
imaginary, the Hamiltonian $H^{(q)}$ is self-adjoint with 
respect to the Klein-Gordon inner product, i.e., the
inner product (11) of Ref.~\cite{p19a}.}

The analysis of the $\Phi_n$ is equivalent to the study of
the eigenvectors of the Laplacian over a three-dimensional
group manifold $\Sigma_t$. The general problem is the subject
of the investigation in spectral geometry which is beyond the
scope of the present article. However, let us  recall some
well-known facts about spectral properties of the Laplacian
$\Delta$ for an arbitrary finite-dimensional Riemannian
manifold $\Sigma$ without boundary.

The following results are valid for the case where
$\Sigma$ is compact or the eigenfunctions are required to
have a compact support\footnote{This is equivalent with the
case where $\Sigma$ has a boundary $\partial\Sigma$, over
which the eigenfunctions vanish.}
\cite{ga-hu-la}:
	\begin{itemize}
	\item[1.] The spectrum of $\Delta$ is an infinite discrete
	subset of  non-negative real numbers.
	\item[2.] The eigenvalues are either non-degenerate or
	finitely degenerate. 
	\item[3.] There is an orthonormal set of eigenfunctions
	which form a basis for $L^2(\Sigma)$.
	\item[4.] If $\Sigma$ is compact,  then the first eigenvalue
	is zero which is non-degenerate with the eigenspace
	given by the set of constant functions, i.e., $\xC$. If 
	$\Sigma$ is not compact but the eigenfunctions are
	required to have a compact support, then the first
	eigenvalue is positive.
	\end{itemize}

Another piece of useful information about the spatially 
homogeneous cosmological models is that (up to a multiple of 
$i=\sqrt{-1}$) the invariant vector fields $\hat X_a$ yield a 
representation of the generators $L_a$ of $G$, with $L^2(\Sigma_t)$ 
being the representation space, one can view the Laplacian $\Delta_t$
as (a representation of ) an element of the enveloping algebra
of the Lie algebra of $G$. Therefore, $\Delta_t$ commutes with
any Casimir operator ${\cal C}_\lambda$ and consequently
shares a set of simultaneous eigenvectors with
${\cal C}_\lambda$. This in turn suggests one to specialize
to particular subrepresentations with definite
${\cal C}_\lambda$. In particular for compact groups, this
leads to a reduction of the problem to a collection of
finite-dimensional ones.\footnote{Here I mean a
finite-dimensional Hilbert space.}

In the remainder of this article I shall  try to employ these
considerations to investigate some specific models.

\section*{IV~~Bianchi Type I}

In this case $G$ is Abelian, therefore $X_a$ are themselves
Casimir operators and the eigenfunctions of  $\Delta_t$, i.e., $\Phi_n$,
are independent of $t$. Hence the Berry connection one-form
(\ref{be-co}) and its non-Abelian generalization (\ref{connection})
vanish identically, and the geometric phase is trivial.

\section*{V~~Bianchi Type IX}

In this case $G=SU(2)=S^3$. The total angular momentum operator
$\hat J^2=\sum_a \hat J_a^2$ is a Casimir operator. Therefore,
I shall consider the subspaces ${\cal H}_j$ of ${\cal H}_t=
L^2(S^3_t)$ of definite angular momentum $j$. The left-invariant
vector fields $\hat X_a$ are given by $\hat X_a=i\hat J_a$, in
terms of which Eq.~(\ref{lie-al}), with $C_{ab}^c=\epsilon_{abc}$,
is written in the familiar form:
	\xbe
	\left[ \hat J_a,\hat J_b\right]=i\epsilon_{abc}\hat J_c\;,
	\label{lie-al-su2}
	\xee
with $\epsilon_{abc}$ denoting the totally antisymmetric Levi
Civita symbol and $\epsilon_{123}=1$. 

Eq.~(\ref{eg-va-eq-la}) takes the form:
	\xbe
	(\hat{H'}+k^2_n)\Phi_n=0\;,
	\label{eg-va-eq-tilde}
	\xee
where $\hat{H'}$ is an induced Hamiltonian defined by
	\xbe
	\hat H':=	g^{ab}(t) \hat J_a\hat J_b\;,
	\label{tilde-h}
	\xee
and $k^2_n:=E_n^2-\mu^2$. Therefore, the problem of the
computation of the geometric phase is identical with
that of the non-relativistic quantum mechanical system
whose Hamiltonian is given by (\ref{tilde-h}). In
particular, for the mixmaster spacetime, i.e., for $g_{ab}$
diagonal, the problem is identical with the quantum mechanical
problem of a non-relativistic asymmetric
rotor, \cite{hu}.

Another well-known non-relativistic
quantum mechanical effect which is described by
a Hamiltonian of the form (\ref{tilde-h}) is the quadratic
interaction of a spin with a variable electric field $(E_a)$.
The interaction potential is the Stark Hamiltonian: $\hat H_S
=\epsilon (\sum_a E_a\hat J_a)^2$. The phenomenon of the
geometric phase for the Stark Hamiltonian for spin $j=3/2$,
which involves Kramers degeneracy \cite{messiyah}, was
first considered by Mead \cite{mead}. Subsequently,
Avron,~et~al~\cite{av-sa-se-si,av2} conducted a thorough
investigation of the traceless quadrupole Hamiltonians of the
form (\ref{tilde-h}). 

The condition on the trace of the Hamiltonian is
physically irrelevant, since the addition of any
multiple of the identity operator to the Hamiltonian
does not have any physical consequences. In general,
one can express the Hamiltonian (\ref{tilde-h}) in the
form $\hat{H'}=\hat{\tilde H'}+\hat H'_0$, where
$\hat{\tilde H'}:=Tr(g^{ab})\hat J^2/3$,
	\xbe
	\hat H'_0[R]:=\sum_{A=1}^5 R^A\: \hat e_A\;,	
	\label{h'0}
	\xee
is the traceless part of the Hamiltonian, and
	\xbea
	\hat e_1&:=&J_3^2-\frac{1}{3}\: \hat J^2\;,~~~~~~~~
	\hat e_2\::=\: \frac{1}{\sqrt{3}}\: \{\hat J_1,\hat J_2\}\;,
	\xnn\\
	\hat e_3&:=&\frac{1}{\sqrt{3}}\: \{\hat J_2,\hat J_3\}
	\;,~~~~~~~~\hat e_4\::=\: \frac{1}{\sqrt{3}}\:(\hat J_1^2
	-\hat J_2^2)\;,
	\label{e's}\\
	\hat e_5&:=& \frac{1}{\sqrt{3}}\:\{\hat J_1,\hat J_2\}\;,
	\xnn\\
	R^1&:=&g^{33}-\frac{1}{2}\:(g^{11}+g^{22})
	\;,~~~~~~~~R^2\::=\:\sqrt{3}\: g^{13}\;,\xnn\\
	R^3&:=&\sqrt{3}\: g^{23}\;,~~~~~~~~
	R^4\::=\: \frac{\sqrt{3}}{2}\:(g^{11}-g^{22})\;,
	\label{R's}\\
	R^5&:=&\sqrt{3}\: g^{12}\;.\xnn
	\xeea
As shown in Refs.~\cite{av-sa-se-si,av2} the space
${\cal M}'$ of all traceless Hamiltonians of  the form
(\ref{h'0}) is a five-dimensional real vector space. The
traceless operators $\hat e_A$ form an orthonormal
\footnote{Orthonormality is defined by the inner product
$\xbr \hat A,\hat B\xkt:=3\:Tr(\hat A\hat B)/2$.} basis for
${\cal M}'$. Removing the point $(R^A=0)$ from ${\cal M}'$,
to avoid the collapse of all eigenvalues, one has 
the space $\xR^5-\{ 0\}$ as the parameter space. The
situation is analogous to Berry's original example of a
magnetic dipole in a changing magnetic field,
\cite{berry1984}. Again, a rescaling of the Hamiltonian
by a non-zero function of $R^A$ does not change the
geometric phase. Thus the relevant parameter space is
${\cal M}=S^4$. Incidentally, the point corresponding
to $0\in\xR^5$ which is to be excluded, corresponds to
the class of Friedmann-Robertson-Walker models.

Following Berry \cite{berry1984} and Ref.~\cite{p19a},
let us identify $t$ with the affine parameter of a 
curve $C:[0,\tau]\to S^4$ traced by the parameters 
$R^A$ in $S^4$. Such a curve may be defined
by the action of the group $SO(5)$ which acts transitively
on $S^4$. Therefore the time-dependence of the
Hamiltonian may be realized by an action of the
group $SO(5)$ on a fixed Hamiltonian. As it is discussed
in Ref.~\cite{av2}, it is the unitary representations
${\cal U}$ of the double cover $Spin(5)=Sp(2)$
of $SO(5)$ (alternatively the projective representations
of $SO(5)$) which define the time-dependent
Hamiltonian:
	\xbe
	\hat H'_0[R(t)]={\cal U}[g(t)]\: \hat H'_0[R(0)]\:
	 {\cal U}[g(t)]^\dagger\;.
	\label{time-dep}
	\xee
Here, $g:[0,\tau]\to Sp(2)$, is defined by $R(t)=:\pi[g(t)]
R(0)$, where $\pi :Sp(2)\to SO(5)$ is the canonical
two-to-one covering projection and $R(t)$ corresponds to the point
$C(t)\in S^4$. The emergence of the
group $Sp(2)$ is an indication of the existence of a
quaternionic description of the system, \cite{av2}.

Let us next examine the situation for irreducible
representations $j$ of $SU(2)$. As I previously
described, $\hat J^2$ commutes with the Hamiltonian.
Hence the Hamiltonian is block-diagonal in the basis
with definite total angular momentum $j$.
For each $j$, the representation space ${\cal H}_j$ is
$2j+1$ dimensional. Therefore the restriction of the
Hamiltonian $\hat{H'}_0$ to  ${\cal H}_j$ and
${\cal U}[g(t)]$ are $(2j+1)\times(2j+1)$ matrices.

Let $\{ \phi^I_{j_n}\}$ be a complete set of orthonormal
eigenvectors of the initial Hamiltonian $\hat H'_0[R(0)]$,
where $I$ is a degeneracy label. Then the eigenvectors
of $\hat H'_0[R]$ are of the form
	\xbe
	\Phi^I_{j_n}[R]={\cal U}[g]\:\phi^I_{j_n}\;,
	\label{phi=u-phi}
	\xee
and the non-Abelian connection one-form
(\ref{connection}) is given by
	\xbe
	{\cal A}_{j_n}^{IJ}=i\xbr \phi^I_{j_n}|\:
	{\cal U}^\dagger
	d\,{\cal U}\:|\phi^J_{j_n}\xkt\;.
	\label{be-co-bc}
	\xee

For the integer $j$, bosonic systems, it is known that
the quadratic Hamiltonians of the form (\ref{h'0}), describe
time-reversal-invariant systems. In this case it can be
shown that the curvature two-form associated with the
Abelian Berry connection one-form (\ref{be-co})
vanishes identically \cite{av2}. The connection
one-form is exact (gauge potential is pure gauge)
and a nontrivial geometric phase can only be
topological, namely it may still exist provided
that the first homology group of the parameter space
is nontrivial.  For the problem under investigation
${\cal M}=S^4$, and the first homology group is
trivial. Hence, in general, the Abelian geometric
phase is trivial. The same conclusion cannot
however be reached for the non-Abelian geometric phases.

In the remainder of this section, I shall examine
the situation for some small values of $j$.
	\begin{itemize}
	\item[1)] $j=0$: The corresponding Hilbert subspace
is one-dimensional. Geometric phases do not arise.
	\item[2)] $j=1/2$: In this case, $\hat J_a=\sigma_a/2$,
where $\sigma_a$ are Pauli matrices. Using the well-known
anticommutation (Clifford algebra) relations $\{\sigma_a,
\sigma_b\}=2 \delta_{ab}$, one can easily show that in this
case
	\xbe
	\hat H'=\frac{1}{2}\sum_{a}g^{aa}(t)\:\hat I\;,
	\label{s=1/2}
	\xee
where $\hat I$ is the identity matrix. Therefore, the
eigenvectors $\Phi_n$ are constant ($g_{ab}$-independent),
the connection one-form (\ref{be-co-bc}) vanishes, and no
nontrivial geometric phases occur.
	\item[3)] $j=1$: In this case the Hilbert subspace is
three-dimensional. The Abelian geometrical phases are
trivial. The nontrivial matrix-valued geometrical phases may be
present, provided that the Hamiltonian has a degenerate
eigenvalue.  Using the ordinary $j=1$ matrix representations
of $\hat J_a$, one can easily express $\hat H'_0$ as a
$3\times 3$ matrix. It can then be checked that in the generic
case the eigenvalues of $\hat H'_0$ are not 
degenerate.\footnote{This is true for all integer $j$,
\cite{av2}.} However, there are cases for which a
degenerate eigenvalue is present. A simple example is the
Taub metric, $(g_{ab})={\rm diag}(g_{11},g_{22},g_{22})$,
which is a particular example of the mixmaster metric, 
\cite{ra-sh}. For the general mixmaster metric, the
eigenvalue problem can be easily solved. The
eigenvalues of the total Hamiltonian $\hat H'$ of
Eq.~(\ref{tilde-h}) are
	\[
	\frac{1}{g_{11}}+\frac{1}{g_{22}}
	\;,~~~~~\frac{1}{g_{11}}+\frac{1}{g_{33}}
	\;,~~~~~\frac{1}{g_{22}}+\frac{1}{g_{33}}\;.\]
Therefore the degenerate case corresponds to the
coincidence of at least two of $g_{aa}$'s, i.e.,
a Taub metric. However, even in the general mixmaster
case, one can find a constant ($g_{aa}$-independent)
basis which diagonalizes the Hamiltonian. Hence
the non-Abelian connection one-form vanishes
and the geometric phase is again trivial. This is not
however the case for general metrics. In the Appendix,
it is shown that without actually solving the general
eigenvalue problem for the general Hamiltonian, one can
find the conditions on the metric which render 
one of the eigenvalues of the Hamiltonian
degenerate.  Here, I summarize the results. Using the
well-known matrix representations of the angular
momentum operators $\hat J_a$ in the $j=1$
representation \cite{schiff} one can write the
Hamiltonian (\ref{tilde-h}) in the form:
	\xbe
	\hat H'=\left(\begin{array}{ccc}
	t+2z&\xi^*&\zeta^*\\
	\xi&2(t+z)&-\xi^*\\
	\zeta&-\xi&t+2z\end{array}\right)\;,
	\label{tilde-h'-0}
	\xee
where
	\xbea
	t&:=&\frac{1}{2}\:(g^{11}+g^{22})-g^{33}\;,~~~~~
	z\::=\:g^{33}\;,\xnn\\
	\xi&:=&\frac{1}{\sqrt{2}}\:(g^{13}+ig^{23})\;,~~~~~
	\zeta\::=\: \frac{1}{2}\:(g^{11}-g^{22})+ig^{12}\;.
	\xnn
	\xeea
Then it can be shown (Appendix) that the necessary
and sufficient conditions for $\hat H'$ to have a
degenerate eigenvalue are 
	\begin{itemize}
	\item[I.] \underline{for $\zeta=0$:}  $\xi=0$, in which case,
$\hat H'$ as given by Eq.~(\ref{tilde-h'-0}) is already
diagonal. The degenerate and non-degenerate eigenvalues
are $t+2z$ and  $2(t+z)$, respectively. In terms of the
components of the metric, these conditions can be written
as: $g_{11}=g_{22}$ and $g_{ab}=0$ if $a\neq b$. This
is a Taub metric which as discussed above does not
lead to a nontrivial  geometric phase.
	\item[II.] \underline{for $\zeta\neq 0$:}  
	\xbe
	\zeta={\cal Z}\,e^{2i\theta}\;,~~~~~ 
	t={\cal Z}-|\xi|^2/{\cal Z}\;, 
	\label{pa-eqs}
	\xee
where $\exp[i\theta]:=\xi/|\xi|$ and  ${\cal Z}\in\xR-\{ 0\}$.
In this case the degenerate and
non-degenerate eigenvalues are $2({\cal Z}+z)-|\xi|^2/
{\cal Z}$ and $2(z-|\xi|^2/{\cal Z})$, respectively.
\end{itemize}
For the latter case, an orthonormal set of eigenvectors
is given by:
	\xbe
	v_1=\frac{1}{\sqrt{2}}\,\left(\begin{array}{c}
	e^{-2i\theta}\\0\\1
	\end{array}\right)\,,~~~~
	v_2=\frac{1}{\sqrt{1+2{\cal X}^2}}\,
	\left(\begin{array}{c}
	{\cal X}e^{-i\theta}\\1\\-{\cal X}e^{i\theta}
	\end{array}\right)\,,~~~~
	v_3=\frac{1}{\sqrt{2(1+2{\cal X}^2)}}\,
	\left(\begin{array}{c}
	-e^{-2i\theta} \\2{\cal X}e^{-i\theta}\\1
	\end{array}\right)\,,
	\label{v's}
	\xee
where ${\cal X}:=|\xi|/(2{\cal Z})$. In view of the general
argument valid for all non-degenerate eigenvalues,
the geometric phase associated with $v_3$ is trivial.
This can be directly checked by substituting
$v_3$ in the formula (\ref{be-co})
for the Berry connection one-form.  This leads, after
some algebra, to the surprisingly simple result ${\cal A}_{33}
:=i\xbr v_3|dv_3\xkt=d\theta$. Therefore, ${\cal A}_{33}$
is exact as expected, and the corresponding geometric
phase is trivial. Similarly one can compute the
matrix elements ${\cal A}_{rs}:=i\xbr v_r|dv_s\xkt$,
$r,s=1,2$, of the non-Abelian connection one-form (\ref{connection}).
The result is
	\xbea
	{\cal A}&=&\left(\begin{array}{cc}
	1&{\cal F}\\
	{\cal F}^*&0
	\end{array}\right)\:\omega\;,
	\label{be-co-v's}\\
	{\cal F}&:=&\frac{2{\cal X}\,e^{i\theta}}{
	\sqrt{2(1+2{\cal X}^2)}}\:=\:
	\frac{2\epsilon\xi}{\sqrt{2+\left|\frac{\xi}{\zeta}
	\right|^2}}\:=\:\frac{\epsilon(g^{13}+ig^{23})}{
	\sqrt{1+\frac{(g^{13})^2+(g^{23})^2}{ (g^{11}-
	g^{22})^2+(2g^{12})^2}}}\;,\xnn\\
	\omega&:=&d\theta\:=\:\frac{g^{13}dg^{23}-
	g^{23}dg^{13}}{(g^{13})^2+(g^{23})^2}\;,\xnn
	\xeea
where $\epsilon:={\cal Z}/|\zeta|=\pm 1$.
As seen from Eq.~(\ref{be-co-v's}), ${\cal A}$ is
a $u(2)$-valued one-form, which vanishes
if  $g^{23}/g^{13}$ is kept constant during the evolution
of the universe. 

It is also worth mentioning that the requirement of
the existence of degeneracy is equivalent to
restricting the parameters of the system to
a two-dimensional subset of $S^4$. Thus, the
corresponding spectral bundle \cite{si,p6} is a $U(2)$
vector bundle over a two-dimensional parameter space
$\tilde{\cal M}$. The manifold structure of $\tilde{\cal M}$
is determined by Eqs.~(\ref{pa-eqs}).  In terms of the
parameters $R^A$ of (\ref{R's}), these equations are
expressed by
	\xbe
	R^5=f_1\,R^4\;,~~~~~R^1=f_2\,R^4+
	\frac{f_3}{R^4}\;,
	\label{r5-r1}
	\xee
where
	\xbea
	f_1&:=&\frac{2R^2R^3}{(R^2)^2-(R^3)^2}\;,~~~~~
	f_2\::=\:\pm\frac{(R^2)^2+(R^3)^2}{\sqrt{3}\,[
	(R^2)^2-(R^3)^2]}\;,\xnn\\
	f_3&:=&\mp\frac{(R^2)^2-(R^3)^2}{2\sqrt{3}}\;,~~~~~
	f_4\::=\:(R^2)^2+(R^3)^2\;.
	\xnn
	\xeea
Here $f_4$ is also introduced for future use. In addition
to (\ref{r5-r1}), one also has the condition $(R^A)\in S^4$.
If $S^4$ is identified with the round sphere, this condition
takes the form $\sum_A (R^A)^2=1$. Substituting
(\ref{r5-r1}) in this equation, one finds
	\xbe
	(1+f_2^2+f_3^2)(R^4)^4-(1-f_4-2f_2f_3)(R^4)^2
	+f_3^2=0\;.
	\label{r4=f'a}
	\xee
Eq.~(\ref{r4=f'a}) may be easily solved for $R^4$. This
yields
	\xbe
	R^4=\pm\,\frac{3[(R^2)^2-(R^3)^2]^2}{8[(R^2)^2+
	(R^3)^2]^2}\:\left[ 1-\frac{2}{3}[(R^2)^2+(R^3)^2]
	\pm\sqrt{1-\frac{4}{3}[(R^2)^2+(R^3)^2]}\:\right]\;.
	\label{r4=}
	\xee
Note that the parameters $R^A$ are related to the
components of the inverse of the three-metric through
Eqs.~(\ref{R's}). Thus the parameter space
$\tilde{\cal M}$ is really a submanifold of the
corresponding minisuperspace.
Fig.~1 shows a three-dimensional plot of $R^4$ as a
function of $R^2$ and $R^3$, i.e., a plot of the
parameter space $\tilde{\cal M}$ as embedded in $\xR^3$.
Note that $R^2=\pm R^3$ renders $f_1$
and $f_2$ singular. The corresponding points which are
depicted as the curves along which the figure becomes
non-differentiable must be handled with care.
The smooth part of $\tilde{\cal M}$ consists of eight connected
components, each of which is diffeomorphic to an open
disk (alternatively $\xR^2$). 
	\item[4)] $j=3/2$: This case has been studied in
Refs.~\cite{av-sa-se-si,av2} in detail. Therefore I suffice
to note that it involves nontrivial geometric phases. Note that
because of Kramer's degeneracy, one does not need to 
restrict the minisuperspace to obtain degenerate eigenvalues.
Every solution of the Bianchi type IX model involves a 
non-Abelian geometric phase. 
	\end{itemize}

\section*{VI~~Conclusion}
In this article I applied the method developed in Ref.~\cite{p19a}
to investigate the existence of cyclic and noncyclic adiabatic geometric
phases induced by spatially homogeneous cosmological backgrounds on
a complex Klein-Gordon field. Unlike the examples presented in 
Ref.~\cite{p19a}, here the freedom in the choice of  the decomposition 
parameter $q$ turned out to simplify the analysis. 

I showed that for the Bianchi type I models Berry's connection one-form
vanished identically. This was not the case for the Bianchi type IX models.
Hence, for these models nontrivial non-Abelian adiabatic geometric phases
could occur in general. A rather interesting observation was the relationship
between the induced Hamiltonians in the Bianchi type IX models and the
quadrupole Hamiltonians of the molecular and nuclear physics. I also 
showed that even for the integer spin representations nontrivial geometric 
phases could exist. This should also be of interest for the molecular 
physicists and chemists who have apparently investigated only the 
fermionic systems  (half-integer spin representations.) A rather thorough
investigation of the non-Abelian adiabatic geometric phase for arbitrary 
spin 1 systems has been conducted in \cite{p20}. 

As described in Ref.~\cite{p19a} the arbitrariness in the choice of $q$ 
leads to a $GL(1,\xC)$ symmetry of the two-component formulation of the
Klein-Gordon equation. In the context of general relativity where the 
Poincar\'e invariance is replaced by the diffeomorphism invariance, one can
use the time-reparameterization symmetry of the background gravitational 
field and the geometric phase to absorb the magnitude $|q|$ of the 
decomposition parameter $q$ into the definition of the lapse function 
$N=(-g^{00})^{-1/2}$. In this way only a $U(1)$ subgroup of the 
corresponding $GL(1,\xC)$ symmetry group survives. The $GL(1,\xC)$ or 
$U(1)$ symmetry associated with the freedom of choice of the 
decomposition parameter seems to have no physical basis or consequences. 
It is merely a mathematical feature of the two-component formalism which
can occasionally be used to simplify the calculations.

The application of the two-component formulation for the Bianchi models 
manifestly shows that this method can be employed even for the cases 
where the background spacetime is non-stationary. One must however 
realize that the present analysis is only valid within the framework of  the 
relativistic adiabatic approximation \cite{p19a}. Although, the 
(approximate) stationarity of  the background  metric is a sufficient 
condition for the validity of the adiabatic approximation, it is not necessary. 
This can be easily seen by noting that for example in the case of Bianchi  
IX model, for spin $j=1/2$ states, one has $\dot\Phi_n=0$, so 
$\xbr \Phi_m|\dot\Phi_n\xkt=0$. Therefore, although the spacetime is not 
stationary, the adiabatic approximation yields the exact solution of the field
equation. This shows that in general for arbitrary non-stationary spacetimes,
there may exist adiabatically evolving states to which the above analysis 
applies. 

The extension of our results to the non-adiabatic cases requires a 
generalization of the analysis of Ref.~\cite{p19a} to non-adiabatic 
evolutions. 

\section*{Acknowledgments}
I would like to thank Bahman Darian for many fruitful
discussions and helping me with the computer graphics, 
Teoman Turgut for bringing to my attention one of the
references, and Kamran Saririan for mailing me copies
of a couple of the references.

\section*{Appendix}
In this Appendix I show how one can obtain the conditions
under which the Hamiltonian (\ref{tilde-h'-0}) has
degenerate eigenvalues without actually solving the
eigenvalue problem in the general case.

The analysis can be slightly simplified if one writes the
Hamiltonian (\ref{tilde-h'-0}) in the form:
	\xbea
	\hat H'&=&(t+2z)\hat I+\hat{\tilde H}\;,\xnn\\
	\hat{\tilde H}&:=&\left(\begin{array}{ccc}
	0&\xi^*&\zeta^*\\
	\xi&t&-\xi^*\\
	\zeta&-\xi&0\end{array}\right)\;,
	\label{tilde-h'}
	\xeea
where $\hat I$ is the $3\times 3$ identity matrix.  Clearly,
the eigenvalue problems for $\hat H'$ and $\hat{\tilde H}$
are equivalent. Computing the characteristic polynomial
for $\hat{\tilde H}$, i.e., $P(\lambda):=\det(\hat{\tilde H}-
\lambda \hat I)$, one finds:
	\xbe
	P(\lambda)=-\lambda^3+t\lambda^2+
	(|\zeta|^2+2|\xi|^2)\lambda-(t|\zeta|^2+\zeta\xi^{*2}
	+\zeta^*\xi^2)\;.
	\label{p}
	\xee
If one of the eigenvalues (roots of $P(\lambda)$) is
degenerate, then
	\xbe
	P(\lambda)=-(\lambda-l_1)(\lambda-
	l_2)^2\;.
	\label{p'}
	\xee
Comparing Eqs.~(\ref{p}) and (\ref{p'}), one finds
	\xbe
	t=l_1+2l_2\;,~~~l_2^2+2l_1l_2=-(|\zeta|^2+
	2|\xi|^2)\;,~~~l_1l_2^2=-(t|\zeta|^2+\zeta\xi^{*2}+
	\zeta^8\xi^2)\;.
	\label{1}
	\xee
Furthermore since $l_2$ is at least doubly degenerate,
the rows of the matrix:
	\xbe
	\hat{\tilde H}-l_2\: \hat I=
	\left(\begin{array}{ccc}
	-l_2&\xi^*&\zeta^*\\
	\xi&t-l_2&-\xi^*\\
	\zeta&-\xi&-l_2\end{array}\right)\;,
	\label{cof}
	\xee
must be mutually linearly dependent. In other words the
cofactors of all the matrix elements must vanish.
Enforcing this condition for the matrix elements
and using Eqs.~(\ref{1}), one finally finds that
either $\xi=\zeta=l_2=0$ and $l_1=t$, or 
	\xbea
	\zeta&=&{\cal Z}\,e^{2i\theta}\;,~~~~~
	 t \:=\:{\cal Z}-\frac{|\xi|^2}{{\cal Z}}\;, \xnn\\
	l_2&=&{\cal Z}\;,~~~~~~l_1\:=\:-({\cal Z}+
	\frac{|\xi|^2}{{\cal Z}})\;,\xnn
	\xeea
where $\exp[i\theta]:=\xi/|\xi|$ and  ${\cal Z}\in\xR-\{ 0\}$.

\newpage
\begin{center}
{\large{\bf Figure Caption:}}
\end{center}
\vspace{.5cm}
	{\bf Figure~1:} This is a plot of 
	$R^4=R^4(R^2,R^3)$. The horizontal plane is the 
	$R^2$-$R^3$-plane and the vertical axis is the 
	$R^4$-axis. The parameter space $\tilde{\cal M}$ 
	is obtained by removing the intersection of this
	figure with the planes defined by: $R^4=0$, $R^2=R^3$
	and $R^2=-R^3$. The intersection involves the curves
	along which the figure becomes non-differentiable.

\newpage

\thispagestyle{empty}
.
\begin{figure}
\epsffile{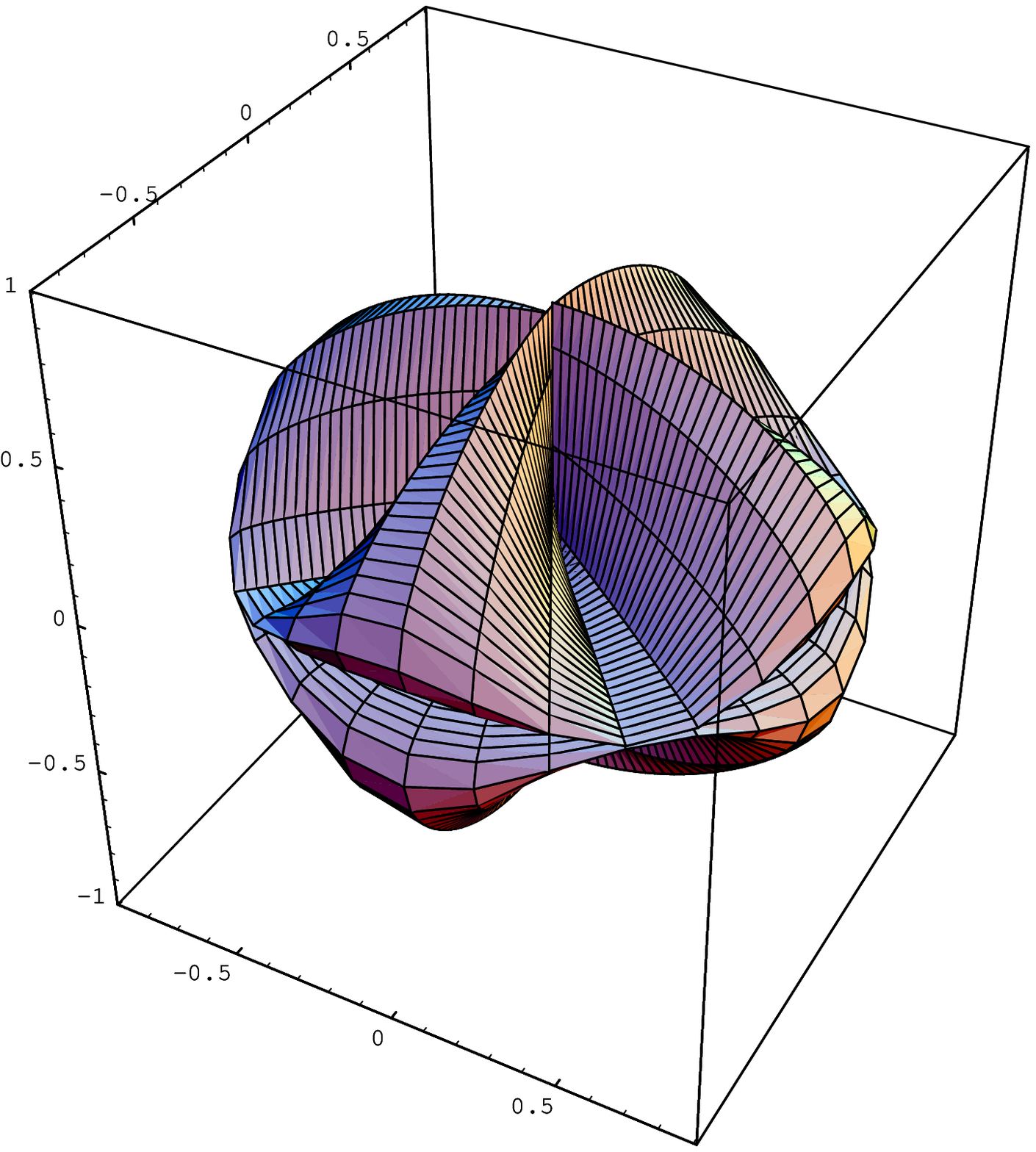}
\end{figure}

\end{document}